\UseRawInputEncoding
\documentclass[notitlepage,12pt]{article}
\usepackage{amsthm}
\usepackage{amsmath}
\usepackage{amssymb}
\usepackage{latexsym}
\usepackage{mathrsfs}
\usepackage{graphicx}
\usepackage{subfigure}
\usepackage{float}
\usepackage{hyperref}
\hypersetup{colorlinks=true, citecolor=red, urlcolor=blue, linkcolor=blue}
\usepackage{color}
\usepackage{mathtools}
\usepackage{dsfont}
\usepackage{indentfirst}
\usepackage{authblk}
\usepackage{cite}
\usepackage{ulem}
\usepackage[thicklines]{cancel}
\usepackage{appendix}
\usepackage{times}
\usepackage[capitalise]{cleveref}

\newtheorem{theorem}{Theorem}
\newtheorem{definition}{Definition}
\newtheorem{lemma}{Lemma}

\newtheorem{corollary}{Corollary}

\newcommand{\ket}[1]{\left|{#1}\right\rangle}
\newcommand{\bra}[1]{\left\langle{#1}\right|}
\newcommand{\braket}[2]{\langle{#1}|{#2}\rangle}
\newcommand{\ketbrad}[1]{\left|{#1}\rangle\!\langle{#1}\right|}
\newcommand{\ketbra}[2]{\left|{#1}\rangle\!\langle{#2}\right|}
\newcommand{\norm}[1]{\left|{#1}\right|}

\begin{document}
\title{A Universal Quantum Certainty Relation for Arbitrary Number of Observables}
\author[1]{Ao-Xiang Liu}
\author[1]{Ma-Cheng Yang}
\author[1,2]{Cong-Feng Qiao\thanks{Corresponding author: \href{mailto:qiaocf@ucas.ac.cn}{qiaocf@ucas.ac.cn}}}
\affil[1]{School of Physical Sciences, University of Chinese Academy of Sciences, Beijing 100049, China}
\affil[2]{Key Laboratory of Vacuum Physics, University of Chinese Academy of Sciences, Beijing 100049, China}
\renewcommand*{\Affilfont}{\small\it} 
\renewcommand*{\Authfont}{\normalsize} 
\renewcommand\Authands{ and } 
\date{}
\maketitle
\vspace{-2em}
	
\begin{abstract}
We derive by lattice theory a universal quantum certainty relation for arbitrary $M$ observables in $N$-dimensional system, which provides a state-independent maximum lower bound on the direct-sum of the probability vectors in terms of majorization relation. While the utmost lower bound coincides with $(1/N,...,1/N)$ for any two observables with orthogonal bases, the majorization certainty relation for $M\geqslant3$ is shown to be nontrivial. The universal majorization bounds for three mutually complementary observables and a more general set of observables in dimension-2 are achieved. It is found that one cannot prepare a quantum state with probability vectors of incompatible observables spreading out arbitrarily. Moreover, we also explore the connections between quantum uncertainty and quantum coherence, and obtain a complementary relation for the quantum coherence as well, which characterizes a trade-off relation of quantum coherence with different bases and is illustrated by an explicit example.
\end{abstract}
	
\section{Introduction}
Heisenberg uncertainty principle~\cite{Heisenberg1927Uber} in quantum mechanics imposes restrictions on the information obtainable in simultaneous measurement of conjugate observables and on the preparation of the quantum states giving definite expectation values for incompatible physical realities. Quantum uncertainty relations have been continuously investigated to characterize the uncertainty principle~\cite{Heisenberg1927Uber}, which is one of the most fundamental features that mark the departure of quantum mechanics from the classical realm. The most representative uncertainty relation is given by Robertson \cite{Robertson1929The}, in which uncertainty has been quantified using the variance of the measurement results. A few other results for variance-based uncertainty relations can be found in Refs. \cite{SE30S,BP14C,KS14H,CB15S,SQ16S,QH16M,SQ17A,YS23M,LA24Q,abdrabbou25}, etc. Over half a century later, it was realized that information entropy allows one to formalize notions like uncertainty and unpredictability \cite{Everett1957Relative,BBI1975U}. Subsequently, the entropic formulation uncertainty relation has been intensely investigated \cite{Deutsch1983Uncertainty, Maassen1988Generalized,WS10E,PJ12U,PJ15S,RA16E,PJ17E}. Studies indicate that these two different forms of uncertainty relations are in fact mutually convertible \cite{Li2016Equivalence}.

Another novel formulation of uncertainty relation is the majorization uncertainty relation. Majorization theory establishes a partial order among probability vectors, providing a characterization of their level of disorder or uncertainty \cite{MW11I}. This framework is inherently well-suited for application to measurement results. The idea to capture uncertainty relations by the majorization approach was proposed by Partovi \cite{Partovi2011Majorization} and later developed by \cite{Friedland2013Universal,Puchala2013Majorization,NV16U,GG18C,Li2019Optimal,XY23Q}. Furthermore, employing the lattice theory on majorization \cite{Cicalese2002Supermodularity}, the optimal quantum uncertainty relation for any number of observables and general measurements has been obtained \cite{Li2019Optimal} and experimentally investigated by \cite{WS22E,WH20E}. The majorization approach is grounded in the intuitive yet remarkably powerful concept that a probability vector formed as a mixture of permutations of another vector exhibits a higher degree of disorder. This makes majorization a more discriminating measure of uncertainty than any measure based on a symmetric, concave function defined on probability vectors, i.e., a Schur-concave function \cite{MW11I}. This property enables the derivation of entropic uncertainty relations from majorization uncertainty relations. Physically, the majorization uncertainty relation implies that it is impossible to prepare a quantum state for which the probability distributions of incompatible observables are simultaneously arbitrarily peaked; in other words, the uncertainties of incompatible observables cannot both be made arbitrarily small.
	
One may ask if quantum theory fundamentally limits the extent to which the uncertainty of incompatible observables can be simultaneously large. The answer is affirmative, i.e., quantum {\it certainty} relation (QCR). Unlike quantum uncertainty relations (of which the sum or product of variances and the sum of the entropies are lower bounded), the QCRs impose nontrivial upper bounds on the variances and the entropies of the measurement outcomes. That is, there is a maximal possible uncertainty, and the joint probability distributions cannot be arbitrarily “spread out”. Hence, the physical meaning of QCRs is that, while QURs prevent us from having perfect knowledge about all incompatible observables,  there is always some residual certainty enforced by quantum theory. Therefore, the QCRs imply another unique feature of preparation uncertainty in quantum theory.

However, in contrast, there is no comparable effort has been made to establish QCRs \cite{PJ17E}. Based on variance measure, Mondal {\it et al}. \cite{Mondal2017Tighter} derived two variance-based QCRs (presented in both product- and sum- form), which put the state-dependent upper bounds on the variances of two incompatible observables. These bounds have been experimentally validated by \cite{XL20D}. Although Korzekwa {\it et al}. \cite{KK14O} generally proved that nontrivial state-independent QCRs do not exist for two arbitrary observables, they do exist, e.g., for more than two measurements described in terms of mutually unbiased bases (MUBs). On the other hand, S\'anchez \cite{Sanchez1993Entropic,Sanchez1995Improved} dealt with the set of observables with their eigenbases being MUBs (we note these observables by MUBs for simplicity hereafter) and obtained an entropic quantum certainty relation for $N+1$ MUBs in $N$-dimensional system, where the existence of a maximal set of $N+1$ MUBs has been proved in the case of the prime power dimension \cite{Durt2010On}, $N=m^{k}$ with $m$ prime and $k$ a positive integer, but is ambiguous for the other composite number dimensions. Recently, Pucha{\l}a {\it et al}. \cite{Puchala2015Certainty} analyzed the upper bounds for the average entropy of orthonomal measurements, and obtained entropic quantum certainty relations that are not restricted to MUBs but apply exclusively to pure states. Very recently, Canturk {\it et al}. \cite{Canturk2021Optimal} derived an entropic quantum certainty relation for $M$ MUBs, which is valid for any state only when $M=N+1$ is satisfied. To our best knowledge, the universal quantum certainty relation, which applies to an arbitrary set of observables and any state, has not been established yet. One of the primary focuses of QCR research is the establishment of universal quantum certainty relations (UQCRs).

In this paper, we propose a universal quantum certainty relation by majorization theory, which is applicable to an arbitrary set of observables and any state. The majorization universal quantum certainty relation shows that one cannot prepare a quantum state with probability vectors of incompatible observables spreading out arbitrarily. It demonstrates that quantum theory puts an intrinsic limit on the extent to which the uncertainty of incompatible observables can be simultaneously large, which is another hallmark of quantum theory. We utilize also the Schur-concave function to establish a universal entropic quantum certainty relation. It is also demonstrated that, using the vectorial measure, the uncertainty is identical to coherence for pure states and derive as well a complementarity relation for quantum coherence using the framework of majorization.

This paper is structured as follows: In Section \ref{QM_ML}, we provide an introduction to the fundamental concepts of quantum measurement and the majorization lattice. The main results are presented in Section \ref{TUCR}. In Section \ref{CR_QS}, we illustrate the UQCR in qubit systems and discuss its advancements. The complementarity relation for quantum coherence is explored in Section \ref{CR_QC}. Finally, Section \ref{Con} gives a conclusion of this work.

\section{Quantum Measurement and Majorization Lattice}
\label{QM_ML}
We start with a brief introduction to quantum measurement formalism and majorization lattice theory.

\subsection{Quantum Measurement}
In $N$-dimensional Hilbert space $\mathcal{H}_N$, a positive operator-valued measurement (POVM), performed on quantum state $\rho$, can be described by a set of measurement operators $\lbrace E_{i}\rbrace$. These operators satisfy the completeness condition $\sum_{i}X_{i}=\mathds{1}$, where $X_{i}=E_{i}^{\dagger}E_{i}$ are referred to as the POVM elements of the measurement \cite{MA2010Quantum}. The probability of obtaining the $i$-th measurement outcome is determined by the Born rule, given as $p_{i}=\mathrm{Tr}[X_{i}\rho]$. In particular, a projective measurement is commonly associated with an observable $X=\sum_{i=1}^{N}x_{i}X_{i}$, where $X_{i}=\ketbra{x_{i}}{x_{i}}$ is the projection operator and $X\ket{x_{i}}=x_{i}\ket{x_{i}}$.
	
In addition, let $\mathcal{M}=\{\ket{m_i}\}$ and $\mathcal{M}^\prime=\{|m^\prime_j\rangle\}$ be two orthonormal bases in $\mathcal{H}_N$. They are called mutually unbiased if and only if for all $i$ and $j$, $|\braket{m_i}{m^\prime_j}|=\frac{1}{\sqrt{N}}$. Consequently, the set $\mathds{M}=\{\mathcal{M}^{(1)},...,\mathcal{M}^{(M)}\}$ forms a set of MUBs if every pair of bases in $\mathds{M}$ are mutually unbiased.

\subsection{Majorization Lattice}
Majorization theory is an elegant tool for exploring the nature of quantum theory and has opened a new horizon of quantum information. For instance, majorization theory has found its significant applications in quantum resource theories \cite{Nielsen1999Conditions,YY15Q,Du15Con,BK16C,Du17E,BG17A,LC19D,liu25}, quantum entanglement detection \cite{OG04E,PM12E,JZ17C,Li2020AnOptimal,Li2021Characterizing,ZG23E,MC23A,yang25w}, and characterizing quantum uncertainty principle \cite{Partovi2011Majorization,Friedland2013Universal,Puchala2013Majorization,NV16U,GG18C,Li2019Optimal,XY23Q}. In this section, in order to make the ensuing study more self-contained, the fundamental definitions and mathematical properties associated with majorization lattice are introduced.

Let us consider a set of probability vectors whose entries are rearranged in non-increasing order, i.e.,
\begin{equation}
\mathcal{P}^{N}=
\Big\lbrace \boldsymbol{p}=
(p_1,\cdots,p_N): p_i\in[0,1],\ \sum_{i=1}^{N}p_i=1,\ p_i\geqslant p_{i+1}\Big\rbrace\ .
\end{equation}
Given any two probability vectors $\boldsymbol{a}$, $\boldsymbol{b}\in\mathcal{P}^{N}$, $\boldsymbol{a}$ is majorized by $\boldsymbol{b}$, in symbols $\boldsymbol{a}\prec\boldsymbol{b}$, if and only if
\begin{equation}
\sum_{i=1}^{k}a_{i}\leqslant\sum_{j=1}^{k}b_{j},\ k\in\lbrace1,...,N\rbrace\ ,
\end{equation}
the equality holds when $k=N$. The physical connotation of majorization relation means that the probability vector $\boldsymbol{a}$ resulting from the measurement $A$ on $\rho$ is said to be less certain than the $\boldsymbol{b}$ concerning the measurement $B$ if $\boldsymbol{a}\prec\boldsymbol{b}$. For a sharp measurement (only one outcome of the measurement occurs), it naturally acquires the probability vectors (1,...,0), which are the maximum probability vectors in terms of majorization order, and only then. Similarly, for the totally uncertain measurements (all possible outcomes occur with equal possibility), it naturally reaches the uniform distribution $(\frac{1}{N},...,\frac{1}{N})$, which is the minimum probability vectors in terms of majorization order. Therefore, the majorization relation is well-suited for characterizing the uncertainty principle. Intuitively, any $\boldsymbol{a}\in\mathcal{P}^{N}$ trivially satisfies $(1/N,\cdots,1/N)\prec\boldsymbol{a}\prec(1,\cdots,0)$. 

Moreover, $\mathcal{P}^N$ equipped with the majorization relation constitutes a lattice \cite{Cicalese2002Supermodularity}:
\begin{definition}[Majorization lattice]
The set $\mathcal{P}^{N}$ endowed with the partial order ``$\prec$" and the operations ``$\vee$" (join), ``$\wedge$" (meet) defines a majorization lattice $\langle \mathcal{P}^{N},\prec,\vee,\wedge\rangle$, that is, for all $\boldsymbol{a}$, $\boldsymbol{b}\in\mathcal{P}^N$, there exist the sumpremum $\boldsymbol{a}\vee\boldsymbol{b}$ and infimum $\boldsymbol{a}\wedge\boldsymbol{b}$ in $\mathcal{P}^{N}$.
\end{definition}

The join operation $\vee$ and the meet operation $\wedge$ find the unique least upper bound and greatest lower bound of two probability vectors, respectively. 
Explicitly, for any two probability vectors $\boldsymbol{a}$, $\boldsymbol{b}\in\mathcal{P}^N$, the components of their meet $\boldsymbol{l}=\boldsymbol{a}\wedge\boldsymbol{b}$ can be determined by:
\begin{equation}
\label{eq:meet components}
l_{i}=\min\big\lbrace \sum_{j=1}^{i}a_{j},\sum_{j=1}^{i}b_{j} \big\rbrace-\min\big\lbrace \sum_{j=1}^{i-1}a_{j},\sum_{j=1}^{i-1}b_{j} \big\rbrace\ .
\end{equation}
Finding the join of two vectors in $\mathcal{P}^{N}$ involves a more intricate process. While we do not explicitly use this operation in what follows, readers interested in the computational details can consult Ref. \cite{Cicalese2002Supermodularity} for the complete algorithm.

In particular, the meet of two probability vectors satisfying the following lemma:
\begin{lemma}
\label{meet_lemma}
For any $\boldsymbol{a},\ \boldsymbol{b}\in\mathcal{P}^{N}$, there exists a greatest lower bound $\boldsymbol{l}=\boldsymbol{a}\wedge\boldsymbol{b}\in\mathcal{P}^{N}$ exhibiting the following properties:
\begin{enumerate}
\item $\boldsymbol{l}\prec\boldsymbol{a}$ and $\boldsymbol{l}\prec\boldsymbol{b}$\ ;
\item  $\forall\ \boldsymbol{c}\in\mathcal{P}^{N}$, if $\boldsymbol{c}\prec\boldsymbol{a}$
	      and $\boldsymbol{c}\prec\boldsymbol{b}$, then $\boldsymbol{c}\prec\boldsymbol{l}$\ .
\end{enumerate}
\end{lemma}

In order to establish UQCR for a given measurement set, we need to determine the infimum of the set of the outcome probability vectors encompassing all quantum states, which is a subset of $\mathcal{P}^N$. Furthermore, it has been proven that the majorization lattice $\mathcal{P}^N$ forms a {\it complete lattice} \cite{Bapat91B, Bondar94I}, which guarantees the existence of the infimum of any subset of itself. Hence, for any subset $\mathcal{S}\subseteq\mathcal{P}^{N}$, there exists the infimum $\boldsymbol{t}\equiv\bigwedge\mathcal{S}$ and supremum $\boldsymbol{s}\equiv\bigvee\mathcal{S}$ of $\mathcal{S}$. As a result, for any set of probability vectors allowed by quantum theory, there exists the infimum and supremum in terms of majorization order. Especially, the components of the infimum $\boldsymbol{t}$ of the subset $\mathcal{S}\subseteq\mathcal{P}^N$ can be determined by \cite{Bosyk19Opt}
\begin{equation}\label{meet_def_set}
t_{i}=\min\big\lbrace\sum_{j=1}^{i}u_{j} | \forall\boldsymbol{u}\in\mathcal{S}\big\rbrace-
\min\big\lbrace\sum_{j=1}^{i-1}v_{j} | \forall\boldsymbol{v}\in\mathcal{S}\big\rbrace,
\ u_{0}=v_{0}=0\ ,
\end{equation}
where $u_{j}$ and $v_{j}$ are the $j$-th components of the probability vectors $\boldsymbol{u}$ and $\boldsymbol{v}$ in $\mathcal{S}$, respectively.
We shall use these properties to achieve our goals in the next section.

\section{The Universal Quantum Certainty Relation}
\label{TUCR}
This section provides an explicit algorithm for computing the infimum of the probability vectors set allowed by quantum theory. 
For a quantum state $\rho$, denote $\lbrace X^{(\alpha)};\boldsymbol{p}^{(\alpha)}\rbrace_{\alpha=1,...,M}$ as a set of $M$ observables $X^{(\alpha)}$ and its corresponding probability vectors $\boldsymbol{p}^{(\alpha)}$. We assemble these vectors by their direct summation, $\boldsymbol{P}=\bigoplus_{\alpha=1}^{M}\boldsymbol{p}^{(\alpha)}$ with $\boldsymbol{p}^{(\alpha)}=(p^{(\alpha)}_{1},...,p^{(\alpha)}_{N})$ , where $p^{(\alpha)}_{i}=\mathrm{Tr}[X^{(\alpha)}_{i}\rho]$ denotes the probability of the $i$-th outcome of the measurement $X^{(\alpha)}$. Note that $\boldsymbol{P}^{\downarrow}\in\mathcal{P}\subseteq\mathcal{P}^{MN}$ is an $MN$-dimensional vector, where the superscript $\downarrow$ indicates that the vector has been rearranged in descending order.
For convenience, all the probability vectors are sorted in descending order, and the superscript $\downarrow$ is omitted hereafter.
\begin{lemma}[Infimum]
\label{lem:inf}
Consider a subset $\mathcal{S}$ of the complete lattice $\mathcal{P}^{N}$, in which the probability vectors are arranged by descending order, then the infimum of $\mathcal{S}$, denoted as $\boldsymbol{t}$, can be computed by
\begin{equation}
\boldsymbol{t}=\boldsymbol{t}^{(1)}\wedge\boldsymbol{t}^{(2)}\wedge\cdots\wedge\boldsymbol{t}^{(N-1)}\ ,
\end{equation}
where $\boldsymbol{t}^{(n)}$ denotes the probability vectors in the subset $\mathcal{S}$ that have the minimum sum of the first $n$ components.
\end{lemma}
\begin{proof}
First to show $\boldsymbol{t}$ is a lower bound. From the definition of $\boldsymbol{t}$ and the associative law of operation $\wedge$ on lattice, we have
\begin{align}
\boldsymbol{t}^{(n)}\prec\boldsymbol{t}\ ,\quad \forall\ n\in{1,...,N-1}\ .
\end{align}
Since $\boldsymbol{t}^{(n)}$ owns the minimum sum of the first $n$ components in set $\mathcal{S}$, $\boldsymbol{t}\prec\boldsymbol{p}$ is satisfied for any $\boldsymbol{p}\in\mathcal{S}$.	
Next, we prove that $\boldsymbol{t}$ is the greatest lower bound. Consider an arbitrary $\boldsymbol{t}'$. If $\boldsymbol{t}'\prec\boldsymbol{p}$ holds for any $\boldsymbol{p}\in\mathcal{S}$, then it must also hold for $\boldsymbol{t}^{(n)}$, $\forall\ n\in{1,...,N-1}$. According to \cref{meet_lemma},
\begin{align}
\label{eq:meet}
\left.
\begin{array}{l}
\boldsymbol{t}'\prec\boldsymbol{t}^{(1)} \\
\boldsymbol{t}'\prec\boldsymbol{t}^{(2)}
\end{array}
\right\} \Rightarrow \boldsymbol{t}'\prec\boldsymbol{t}^{(1)} \wedge \boldsymbol{t}^{(2)}\ .
\end{align}
By repeatedly applying \cref{eq:meet} to $\boldsymbol{t}^{(n)}$, we ultimately obtain $\boldsymbol{t}'\prec\boldsymbol{t}$. Therefore, $\boldsymbol{t}=\boldsymbol{t}^{(1)}\wedge\boldsymbol{t}^{(2)}\wedge\cdots\wedge\boldsymbol{t}^{(N-1)}$ is the infimum of the set $\mathcal{S}$.
\end{proof}

To understand this lemma, one can consider the set $\mathcal{S}$ containing two probability vectors $\boldsymbol{a} = (0.7, 0.2, 0.1)$ and $\boldsymbol{b} = (0.5, 0.45, 0.05)$ in $\mathcal{P}^3$ as an example. For $n=1$, $\boldsymbol{t}^{(1)}$ minimizes the first component while being majorized by both $\boldsymbol{a}$ and $\boldsymbol{b}$. Since the first component of $\boldsymbol{b}$ is smaller (0.5 vs 0.7), $\boldsymbol{t}^{(1)}=\boldsymbol{b}$ has its first component equal to 0.5\ . Similarly, the sum of the first two components is 0.9 for $\boldsymbol{a}$ and 0.95 for $\boldsymbol{b}$, so $\boldsymbol{t}^{(2)}=\boldsymbol{a}$ has the minimum sum of the first two components equal to 0.9\ . From \cref{eq:meet components}, the infimum of $\mathcal{S}$ is $\boldsymbol{t}=\boldsymbol{t}^{(1)}\wedge\boldsymbol{t}^{(2)}=(0.5, 0.4,1)$\ .

To proceed, let us consider the subset $\mathcal{P}=\lbrace\boldsymbol{P}\rbrace\subset\mathcal{P}^{MN}$ encompassing all possible probability vectors for all quantum states, then it follows that the infimum of $\mathcal{P}$ exists by the property of complete lattice. From \cref{lem:inf}, we have
\begin{theorem}[UQCR]
\label{Thm:UCR}
In an N-dimensional quantum system, for an arbitrary set of $M$ observables $\lbrace X^{(\alpha)}\rbrace_{\alpha=1}^{M}$ and quantum state $\rho$, the probability vectors $\boldsymbol{p}^{(\alpha)}$ corresponding to measurements $X^{(\alpha)}$, satisfy the following relation:
\begin{equation}\label{glb_maj}
\boldsymbol{t}\prec\boldsymbol{P}=\bigoplus_{\alpha=1}^{M}\boldsymbol{p}^{(\alpha)}\ .
\end{equation}
Here, $\boldsymbol{t}\coloneqq \boldsymbol{t}^{(1)}\wedge\boldsymbol{t}^{(2)}\wedge\cdots\wedge\boldsymbol{t}^{(MN-1)}$ represents the greatest lower bound (infimum) of $\bigoplus_{\alpha=1}^{M}\boldsymbol{p}^{(\alpha)}$ over all quantum states.
\end{theorem}

Generally, in order to obtain the infimum of the subset $\mathcal{P}$, one needs to calculate the sum of the first $n$ components of $\boldsymbol{t}^{(n)}$ for $n=1,...,N-1$.
Subsequently, we introduce the method to calculate the $\boldsymbol{t}^{(n)}$ of the quantum measurement outcome probability vectors. Let us consider a set of $M$ observables $\lbrace X^{(\alpha)};\boldsymbol{p}^{(\alpha)}\rbrace_{\alpha=1,...,M}$ for quantum state $\rho$ in $N$-dimensional Hilbert space $\mathcal{H}_N$, then we define a set of operators in terms of the projectors $\{|x_i^{(\alpha)}\rangle\langle x_i^{(\alpha)}|\}$ of $X^{(\alpha)}$ as follows:
\begin{equation}\label{set_op}
\mathcal{T}_{\alpha}\coloneqq\Big\lbrace \boldsymbol{x}_{\alpha}|\boldsymbol{x}_{\alpha}=
\sum_{i\in\mathcal{I}_{\alpha}}|x_i^{(\alpha)}\rangle\langle x_i^{(\alpha)}|,\ \mathcal{I}_{\alpha}\subseteq\lbrace 1,\cdots,
N\rbrace\Big\rbrace\ .
\end{equation}
Here, the $|\mathcal{I}_{\alpha}|$ denotes the cardinality of the set $\mathcal{I}_{\alpha}$, and $\sum_{\alpha=1}^{M}|\mathcal{I}_{\alpha}|=n$. 
The operator $\boldsymbol{x}_{\alpha}$ is composed of $|\mathcal{I}_{\alpha}|$ distinct elements sampled from the set $\{|x_i^{(\alpha)}\rangle\langle x_i^{(\alpha)}|\}$.
According to the Born rule, the partial sum of the $|\mathcal{I}_{\alpha}|$ components of probability vector $\boldsymbol{p}^{(\alpha)}$ can be expressed as 
\begin{align}
\sum_{i\in\mathcal{I}_{\alpha}}p_{i}^{(\alpha)}=\mathrm{Tr}[\rho\sum_{i\in\mathcal{I}_{\alpha}}|x_i^{(\alpha)}\rangle\langle x_i^{(\alpha)}|]=\mathrm{Tr}[\rho\ \boldsymbol{x}_{\alpha}]\ .
\end{align}
Thus, the partial sum of the $n$ components of probability vector $\boldsymbol{P}=\bigoplus_{\alpha=1}^{M}\boldsymbol{p}^{(\alpha)}$ (not descendingly rearranged) can be obtained by 
\begin{align}
\gamma_n\coloneqq\sum_{\alpha=1}^{M}\sum_{i\in\mathcal{I}_{\alpha}}p_{i}^{(\alpha)}=\sum_{\alpha=1}^{M}\mathrm{Tr}[\rho\ \boldsymbol{x}_{\alpha}]\ ,\ \sum_{\alpha=1}^{M}|\mathcal{I}_{\alpha}|=n\ ,
\end{align}
which varies with $\boldsymbol{x}_{\alpha}\in \mathcal{T}_\alpha$ corresponding to different sampling according to $\mathcal{I}_\alpha$. 

For a given quantum state $\rho$, we rearrange $\boldsymbol{P}$ by descending order, i.e., $\boldsymbol{P}^\downarrow$ satisfying $P^\downarrow_i\geqslant P^\downarrow_{i+1}$. Hence, the sum of the first $n$ components of $\boldsymbol{P}^\downarrow$ can be obtained by maximizing $\gamma_n$ over different sampling:
\begin{align}
\label{eq:parsum}
\sum_{i=1}^{n}P_{i}=\max\limits_{\substack{\{\mathcal{I}_{\alpha}\} \\ \alpha=1,...,M}}\gamma_n\ .
\end{align}
From the definition of $\boldsymbol{t}^{(n)}$ in \cref{lem:inf}, $\boldsymbol{t}^{(n)}$ denotes the probability vectors that own the minimum sum of the first $n$ components of all possible $\boldsymbol{P}^\downarrow$ allowed by quantum theory, that is,
\begin{align}
\label{min_sum_n}
\Gamma_n\coloneqq\sum_{i=1}^{n}t^{(n)}_i=\min\limits_{\lbrace\rho\rbrace}\max\limits_{\substack{\{\mathcal{I}_{\alpha}\} \\ \alpha=1,...,M}}\gamma_n\ .
\end{align}
The quantum state $\rho_n$ that attains the minimum in \cref{min_sum_n} yields a corresponding $\boldsymbol{t}^{(n)}$.
By \cref{eq:meet components} and \cref{lem:inf}, one can obtain 
\begin{align}
\boldsymbol{t}=(\Gamma_1,\Gamma_2-\Gamma_1,\cdots,\Gamma_{N-1}-\Gamma_{N-2},1-\Gamma_{N-1})\ .
\end{align}

For simplicity, we take qubit system as an example to illustrate the abovementioned procedure. Let $\mathcal{H}_2$ be a two-dimensional Hilbert space, and consider a set of $M$ observables $X^{(\alpha)}$, $\alpha=1,2,\cdots,M$. The density operator of qubit state $\rho$ and observable $X^{(\alpha)}$ can be expressed in terms of the Bloch vectors as $\rho=\frac{1}{2}\left(\mathds{1}+\boldsymbol{r}\cdot\boldsymbol{\sigma}\right)\ ,\ X^{(\alpha)}=\boldsymbol{n}_\alpha\cdot\boldsymbol{\sigma}$,
where $\boldsymbol{r}$ is the Bloch vector, $\boldsymbol{\sigma}=(\sigma_x,\sigma_y,\sigma_z)$ are the Pauli matrices, $\boldsymbol{n}_\alpha$ is the Bloch vector of the observable $X^{(\alpha)}$, and $\mathds{1}$ is the identity operator. The eigenbasis of  $X^{(\alpha)}$ can be written using Bloch vector as $|x_1^{(\alpha)}\rangle\langle x_1^{(\alpha)}|=\frac{1}{2}\left(\mathds{1}+\boldsymbol{n}_\alpha\cdot\boldsymbol{\sigma}\right)\ ,\ |x_2^{(\alpha)}\rangle\langle x_2^{(\alpha)}|=\frac{1}{2}\left(\mathds{1}-\boldsymbol{n}_\alpha\cdot\boldsymbol{\sigma}\right).$
Thus, the index set $\mathcal{I}_\alpha$ can be chosen as $\{1\}$, $\{2\}$, or $\{1,2\}$.
From \cref{set_op}, we have 
\begin{align}
\boldsymbol{x}_\alpha=|x_{1,2}^{(\alpha)}\rangle\langle x_{1,2}^{(\alpha)}|\quad ,\ \text{or}\quad \boldsymbol{x}_\alpha=|x_1^{(\alpha)}\rangle\langle x_1^{(\alpha)}|+|x_2^{(\alpha)}\rangle\langle x_2^{(\alpha)}|=\mathds{1}\ .
\end{align}
The partial sum of the $n$ components of probablity vector $\boldsymbol{P}=\bigoplus_{\alpha=1}^{M}\boldsymbol{p}^{(\alpha)}$ can be expressed as $\gamma_n=\sum_{\alpha=1}^{M}\mathrm{Tr}[\rho\ \boldsymbol{x}_{\alpha}]$.
According to the \cref{min_sum_n}, one can calculate the sum of the first $n$ components of $\boldsymbol{t}^{(n)}$, i.e.,
\begin{align}
\Gamma_n=\min\limits_{\lbrace\rho\rbrace}\max\limits_{\substack{\{\mathcal{I}_{\alpha}\} \\ \alpha=1,...,M}}\gamma_n=\min\limits_{\lbrace\rho\rbrace}\max\limits_{\substack{\{\mathcal{I}_{\alpha}\} \\ \alpha=1,...,M}}\mathrm{Tr}[\rho\ \boldsymbol{x}_{\alpha}]\ ,\ \sum_{\alpha=1}^{M}|\mathcal{I}_\alpha|=n\ .
\end{align}
For example, $\Gamma_1$ has a more simple form in qubit system:
\begin{align}
\Gamma_1=\min\limits_{\lbrace\boldsymbol{r}\rbrace}\max\lbrace \frac{1}{2}+\frac{1}{2}|\boldsymbol{r}\cdot\boldsymbol{n}_{1}|,\frac{1}{2}+\frac{1}{2}|\boldsymbol{r}\cdot\boldsymbol{n}_{2}|,\cdots,\frac{1}{2}+\frac{1}{2}|\boldsymbol{r}\cdot\boldsymbol{n}_{M}| \rbrace\ .
\end{align}
Similarly, calculating $\Gamma_n$ for $n\geq2$ is straightforward and can be effectively implemented through programming for general observables. More specific examples see Examples 1-3 in \cref{CR_QS}.
In summary, we have established a method to compute the infimum of the set of probability vectors allowed by quantum theory. 
	
In \cref{Thm:UCR}, no additional constraints are imposed on the observables in advance. Besides, Eq. (\ref{glb_maj}) also demonstrates optimality for both pure and mixed states. As the counterpart of $\boldsymbol{t}$, the supremum of $\mathcal{P}$, says $\boldsymbol{s}\coloneqq \boldsymbol{s}^{(1)}\vee\boldsymbol{s}^{(2)}\vee\cdots\vee\boldsymbol{s}^{(MN-1)}$, gives the least upper bound of uncertainty relation \cite{Li2019Optimal}, where $\boldsymbol{s}^{(n)}\in\mathcal{P}$ is the probability vector that has the maximum sum of the first $n$ components. The majorization relation can be visualized through the Lorenz curves, initially introduced to compare which is ``more equal" between different income distributions~\cite{Lorenz05M}. Given probability vector $\boldsymbol{P}$, the Lorenz curve is defined as
\begin{equation}
\label{Lorenz curve}
L_{n}(\boldsymbol{P})\coloneqq\sum_{i=1}^{n}P_{i}\ ,\quad L_{0}(\boldsymbol{P})=0\ .
\end{equation}
The Lorenz curves of $\boldsymbol{t}$ and $\boldsymbol{s}$ envelop the curves of all possible probability vectors, i.e.,  $\boldsymbol{t}\prec\ \boldsymbol{P}=\bigoplus_{\alpha=1}^{M}\boldsymbol{p}^{(\alpha)}\prec\boldsymbol{s}$.

One can also derive a series of universal entropic quantum certainty relations by applying Schur-concave (convex) functions~\cite{MW11I}. For instance, the known Shannon entropy is a key concept of information theory and defined by $H(\boldsymbol{p})\coloneqq-\sum_{i}p_{i}\log p_{i},\ p_{i}\geqslant 0,\,\sum_{i} p_{i}=1$. It is Schur-concave so that if $\boldsymbol{p}\prec\boldsymbol{q}$, then $H(\boldsymbol{p})\geqslant H(\boldsymbol{q})$. Naturally, we derive the following universal entropic quantum certainty relation:
\begin{corollary}
\label{Shannon_entropy_relation}
For observables $X^{(\alpha)}$, $\alpha=1,\cdots,M$ and quantum state $\rho$, there exists the following quantum entropic certainty relation
\begin{equation}
\label{Shannon}
\sum_{\alpha=1}^{M}H(X^{(\alpha)})\leqslant H(\boldsymbol{t}\,)\ ,
\end{equation}
where $H(X^{(\alpha)})\coloneqq H(\boldsymbol{p}^{(\alpha)})$ signifies the Shannon entropy of the $\alpha$-th observable $X^{(\alpha)}$ with $\boldsymbol{p}^{(\alpha)}$ the probability vectors of measurements $X^{(\alpha)}$.
\end{corollary}
	
There is an important improvement of the Corollary \ref{Shannon_entropy_relation} by an extra term according to the Theorem 3 of Ref.~\cite{HS10O}:
\begin{equation}\label{Shannon_tighter}
\sum_{\alpha=1}^{M}H(X^{(\alpha)})\leqslant H(\boldsymbol{t}\,)-D(\boldsymbol{P}\Vert\boldsymbol{t})\ ,
\end{equation}
where $D(\boldsymbol{P}\Vert\boldsymbol{t})\equiv\sum_{i}t_{i}\log(\frac{t_{i}}{P_{i}})$ is nonnegative so that the inequality (\ref{Shannon_tighter}) is tighter than (\ref{Shannon}). 
While the bound of the entropic certainty relation for the three Pauli operators $\sigma_i$, $i=x$, $y$, $z$, in Ref. \cite{Sanchez1993Entropic} has long been considered optimal \cite{Sanchez1993Entropic, Puchala2015Certainty, Canturk2021Optimal}, we find that the bound in Eq. (\ref{Shannon_tighter}) is partially tighter than the former (see Example 3 and Figure \ref{Fig-EQCR} in Section \ref{CR_QS}).

\section{UQCRs and Lorenz Curves for Qubit System}
\label{CR_QS}
As illustrations of \cref{Thm:UCR}, we calculate the greatest lower bounds of the uncertainties of two measurements and three measurements in qubit system, and visualize the UQCRs by Lorenz curves. To see the uncertainty region enveloped by majorization lattice, the Lorenz curves of the least upper bounds of the uncertainties are also plotted, which have been presented in \cite{Li2019Optimal}.
	
\begin{figure}[ht]
\centering
\includegraphics[width=0.8\linewidth]{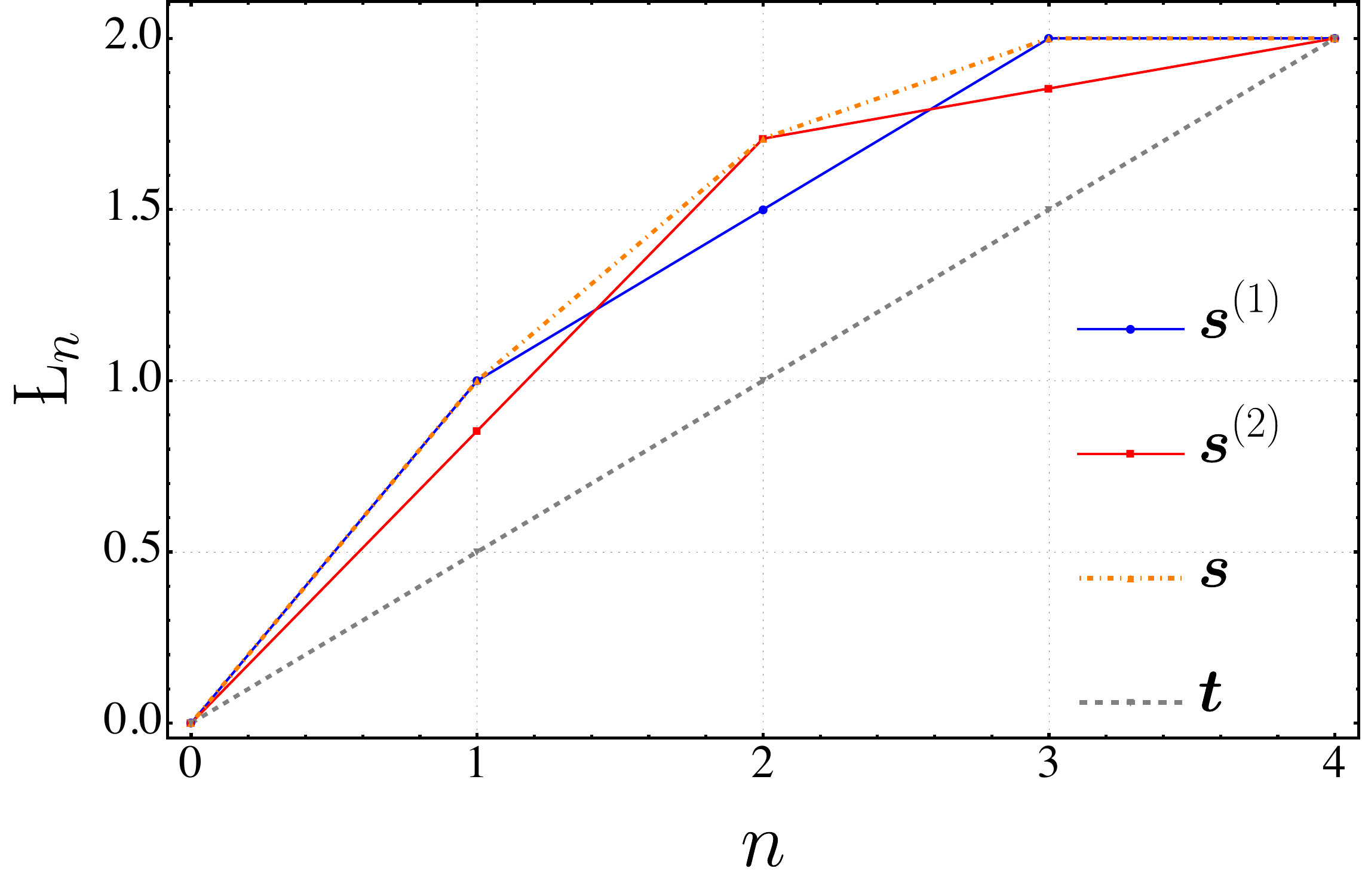}
\caption{The Lorenz curves of the QCR of two observables $X=\sigma_x$ and $Z=\sigma_z$, where $L_{n}$ is the sum of the first $n$ components defined in equation (\ref{Lorenz curve}). By means of $\boldsymbol{t}=\boldsymbol{t}^{(1)}\wedge\boldsymbol{t}^{(2)}$ and $\boldsymbol{s}=\boldsymbol{s}^{(1)}\vee\boldsymbol{s}^{(2)}$ (the least majorization upper bound of $\boldsymbol{P}=\boldsymbol{p}^{(x)}\oplus\boldsymbol{p}^{(z)}$), the Lorenz curves of the $\boldsymbol{t}$ (the gray dashed antidiagonal line) and $\boldsymbol{s}$ (the orange dot-dashed line) respectively give the greatest and the least possible envelope enclosing the curves of $\boldsymbol{P}=\boldsymbol{p}^{(x)}\oplus\boldsymbol{p}^{(z)}$ for all quantum states.}
\label{fig:2o}
\end{figure}

\textbf{Example 1}: Let us consider two observables $X=\boldsymbol{n}_{1}\cdot\boldsymbol{\sigma}$ and $Z=\boldsymbol{n}_{2}\cdot\boldsymbol{\sigma}$ in qubit system, where $\boldsymbol{n}_{1}$ and $\boldsymbol{n}_{2}$ are unit vectors on the Bloch sphere, and $\boldsymbol{\sigma}$ represents the vector composed of the three Pauli matrices. The density matrices of general qubit states can be represented in the Bloch representation as $\rho=\frac{\mathds{1}}{2}+\frac{1}{2}\,\boldsymbol{r}\cdot\boldsymbol{\sigma}$. The probability vectors of measurements $X$ and $Z$ are given by, respectively,
\begin{equation}
\boldsymbol{p}^{(x)}=(\frac{1}{2}+\frac{1}{2}\boldsymbol{r}\cdot\boldsymbol{n}_{1},\frac{1}{2}-\frac{1}{2}\boldsymbol{r}\cdot\boldsymbol{n}_{1}),\ \boldsymbol{p}^{(z)}=(\frac{1}{2}+\frac{1}{2}\boldsymbol{r}\cdot\boldsymbol{n}_{2},\frac{1}{2}-\frac{1}{2}\boldsymbol{r}\cdot\boldsymbol{n}_{2})\; .
\end{equation}
 From \cref{min_sum_n}, the $\Gamma_n$ with $n=1,2,3$ can be calculated as follows:
\begin{align}
\Gamma_1=\min\limits_{\lbrace \boldsymbol{r}\rbrace}\max\limits_{\boldsymbol{n}_1,\boldsymbol{n}_2}\mathrm{Tr}[\rho\ \boldsymbol{x}_{1}]=\min\limits_{\lbrace \boldsymbol{r}\rbrace}\max\lbrace\frac{1}{2}+\frac{1}{2}|\boldsymbol{r}\cdot\boldsymbol{n}_{1}|,\frac{1}{2}+\frac{1}{2}|\boldsymbol{r}\cdot\boldsymbol{n}_{2}|\rbrace\ .
\end{align}
\begin{align}
\Gamma_2=\min\limits_{\lbrace \boldsymbol{r}\rbrace}\max\limits_{\boldsymbol{n}_1,\boldsymbol{n}_2}\mathrm{Tr}[\rho\ \boldsymbol{x}_{2}]=\min\limits_{\lbrace \boldsymbol{r}\rbrace}\max\lbrace & 1,1+\frac{1}{2}\boldsymbol{r}\cdot\boldsymbol{n}_{1}+\frac{1}{2}\boldsymbol{r}\cdot\boldsymbol{n}_{2},1+\frac{1}{2}\boldsymbol{r}\cdot\boldsymbol{n}_{1}-\frac{1}{2}\boldsymbol{r}\cdot\boldsymbol{n}_{2},\notag\\
&1-\frac{1}{2}\boldsymbol{r}\cdot\boldsymbol{n}_{1}+\frac{1}{2}\boldsymbol{r}\cdot\boldsymbol{n}_{2},1-\frac{1}{2}\boldsymbol{r}\cdot\boldsymbol{n}_{1}-\frac{1}{2}\boldsymbol{r}\cdot\boldsymbol{n}_{2}\rbrace\ .
\end{align}
\begin{align}
\Gamma_3=\min\limits_{\lbrace \boldsymbol{r}\rbrace}\max\limits_{\boldsymbol{n}_1,\boldsymbol{n}_2}\mathrm{Tr}[\rho\ \boldsymbol{x}_{3}]=\min\limits_{\lbrace \boldsymbol{r}\rbrace}\max\lbrace \frac{3}{2}+\frac{1}{2}\boldsymbol{r}\cdot\boldsymbol{n}_{1},\frac{3}{2}-\frac{1}{2}\boldsymbol{r}\cdot\boldsymbol{n}_{1},\frac{3}{2}+\frac{1}{2}\boldsymbol{r}\cdot\boldsymbol{n}_{2},\frac{3}{2}-\frac{1}{2}\boldsymbol{r}\cdot\boldsymbol{n}_{2}\rbrace\ .
\end{align}
Furthermore, it is always possible to find a vector on the Bloch sphere perpendicular to both $\boldsymbol{n}_{1}$ and $\boldsymbol{n}_{2}$, in which case $\boldsymbol{t}^{(1)}=(1/2,1/2,1/2,1/2)$ must hold. Similarly, it holds that $\boldsymbol{t}^{(2)}=\boldsymbol{t}^{(3)}=(1/2,1/2,1/2,1/2)$. By \cref{lem:inf}, one can obtain that the greatest lower bound of $\boldsymbol{P}=\boldsymbol{p}^{(x)}\oplus\boldsymbol{p}^{(z)}$ is trivial, i.e., $\boldsymbol{t}=\boldsymbol{t}^{(1)}\wedge\boldsymbol{t}^{(2)}\wedge \boldsymbol{t}^{(3)}=(1/2,1/2,1/2,1/2)$ (See Figure \ref{fig:2o}), which illustrates that nontrivial state-independent QCRs do not exist for two arbitrary observables \cite{KK14O}. 

\textbf{Example 2}: Let us consider three observables $A=1/2\sigma_{x}+1/2\sigma_{y}$, $B=\sigma_{y}$, and $C=\sigma_{z}$, of which the eigenbases are not mutually unbiased. The eigenstates of $A$, $B$, and $C$ on the Bloch sphere are as follows:
\begin{equation}
|x^{\prime}_{\pm}\rangle\langle x^{\prime}_{\pm}|=
\frac{\mathds{1}}{2}\pm\frac{1}{2}\boldsymbol{n}_{x^{\prime}}\cdot\boldsymbol{\sigma}=
\frac{\mathds{1}}{2}\pm\frac{1}{2}\,(\cos\phi\,\boldsymbol{n}_{x}+\sin\phi\,\boldsymbol{n}_{y})\cdot\boldsymbol{\sigma}\ ,
\end{equation}
\begin{equation}
\ketbra{y_{\pm}}{y_{\pm}}=\frac{\mathds{1}}{2}\pm\frac{1}{2}\,\boldsymbol{n}_{y}\cdot\boldsymbol{\sigma}\ ,
\end{equation}
\begin{equation}
\ketbra{z_{\pm}}{z_{\pm}}=\frac{\mathds{1}}{2}\pm\frac{1}{2}\,\boldsymbol{n}_{z}\cdot\boldsymbol{\sigma}\ .
\end{equation}
Here, $\phi=\pi/4$, $\boldsymbol{n}_{x}=(1,0,0)$, $\boldsymbol{n}_{y}=(0,1,0)$, $\boldsymbol{n}_{z}=(0,0,1)$.
\begin{figure}[ht]
\centering  
\includegraphics[width=1\linewidth]{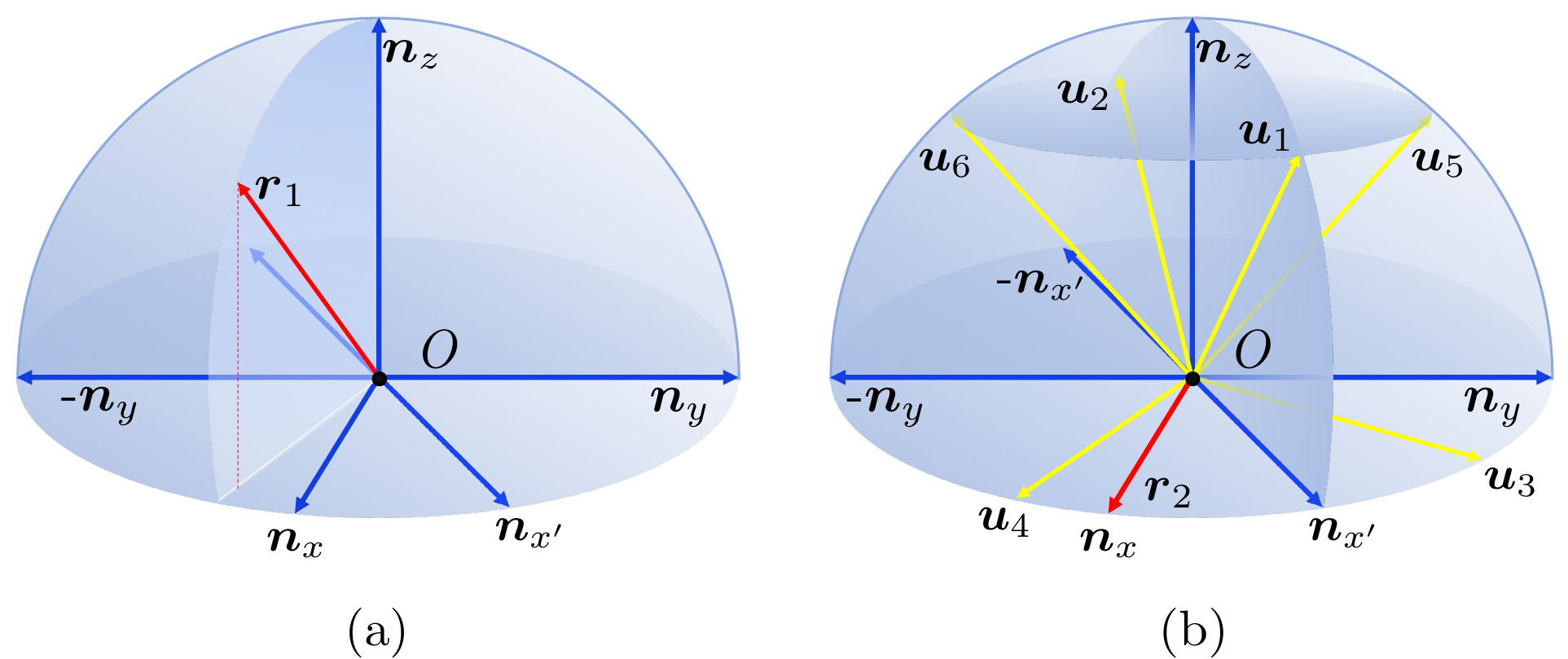}
\vspace{-1em}
\caption{(a) The blue vectors symbolize the Bloch vector of $\boldsymbol{n}_{x,x^{\prime},y,z}$ respectively and the red vector stands for the Bloch vector $\boldsymbol{r}_1$ in Eq. (\ref{Bloch_vec_r1}), which gives $\boldsymbol{t}^{(1)}$. (b) The yellow vectors $\boldsymbol{u}_1$, $\boldsymbol{u}_2$, $\boldsymbol{u}_3$, $\boldsymbol{u}_4$, $\boldsymbol{u}_5$, $\boldsymbol{u}_6$ symbolize $\frac{\boldsymbol{n}_{x^{\prime}}+\boldsymbol{n}_z}{\norm{\boldsymbol{n}_{x^{\prime}}+\boldsymbol{n}_z}}$, $\frac{\boldsymbol{n}_{z}-\boldsymbol{n}_{x^{\prime}}}{\norm{\boldsymbol{n}_{z}-\boldsymbol{n}_{x^{\prime}}}}$, $\frac{\boldsymbol{n}_{x^{\prime}}+\boldsymbol{n}_y}{\norm{\boldsymbol{n}_{x^{\prime}}+\boldsymbol{n}_y}}$, $\frac{\boldsymbol{n}_{x^{\prime}}-\boldsymbol{n}_y}{\norm{\boldsymbol{n}_{x^{\prime}}-\boldsymbol{n}_y}}$, $\frac{\boldsymbol{n}_{y}+\boldsymbol{n}_z}{\norm{\boldsymbol{n}_{y}+\boldsymbol{n}_z}}$, $\frac{\boldsymbol{n}_{z}-\boldsymbol{n}_y}{\norm{\boldsymbol{n}_{z}-\boldsymbol{n}_y}}$ respectively and the red vector is the Bloch vector $\boldsymbol{r}_2=(1,0,0)$, which gives $\boldsymbol{t}^{(2)}$. }
\label{fig:2t}
\end{figure}

We relabel the six eigenstates by $O_i,\,i=1,...,6$ and denote the Bloch vector of $\rho$ as $\boldsymbol{r}=\norm{\boldsymbol{r}}(\sin\theta\cos\xi,\sin\theta\sin\xi,\cos\theta)$, where $\norm{\boldsymbol{r}}$ represents the Euclidean norm of $\boldsymbol{r}$. Therefore, the probabilities can be computed, $\mathrm{Tr}[O_{\pm}\rho]=1/2\pm 1/2\,\boldsymbol{r}\cdot\boldsymbol{n}$. We denote the probability vectors of the $A,B,C$ as $\boldsymbol{p}^{(a)},\boldsymbol{p}^{(b)}$ and $\boldsymbol{p}^{(c)}$ respectively, and calculate the infimum of all the probability vectors $\boldsymbol{P}=\boldsymbol{p}^{(a)}\oplus\boldsymbol{p}^{(b)}\oplus\boldsymbol{p}^{(c)}$. According to Equation (\ref{min_sum_n}), the minimum first component of the vector $\boldsymbol{t}$ can be obtained. Noticing $\boldsymbol{n}_{x^{\prime}}\cdot\boldsymbol{n}_{y}=3\pi/4\geqslant\pi/2$, then $\boldsymbol{t}^{(1)}$ can be obtained by searching for a Bloch vector, say $\boldsymbol{r}_1$, within the solid angle formed by $-\boldsymbol{n}_y$, $\boldsymbol{n}_{x^\prime}$, and $\boldsymbol{n}_z$ such that $\boldsymbol{r}_{1}\cdot\boldsymbol{n}_{x^{\prime},y,z}$ are simultaneously as close to zero as possible. The calculation can be simplified due to the symmetry of the system, i.e., the Bloch vector $\boldsymbol{r}_{1}$ must give the same value for $\boldsymbol{r}_{1}\cdot \boldsymbol{n}_{x^{\prime}}$, $-\boldsymbol{r}_{1}\cdot \boldsymbol{n}_{y}$, and $\boldsymbol{r}_{1}\cdot \boldsymbol{n}_{z}$ (as shown by the red vector in Figure \ref{fig:2t}(a)). Consequently, we obtain and denote $\boldsymbol{r}_{1}$ as
\begin{equation}
\label{Bloch_vec_r1}
\boldsymbol{r}_{1}=\norm{\boldsymbol{r}}(\sin\vartheta\cos\varphi,\sin\vartheta\sin\varphi,\cos\vartheta)\ ,
\end{equation}
where $\tan \vartheta=\frac{1}{\sin\varphi}$, $\varphi=\frac{\pi}{4}-\frac{\phi}{2}$. 

\begin{figure}[ht]
\centering
\includegraphics[width=1\linewidth]{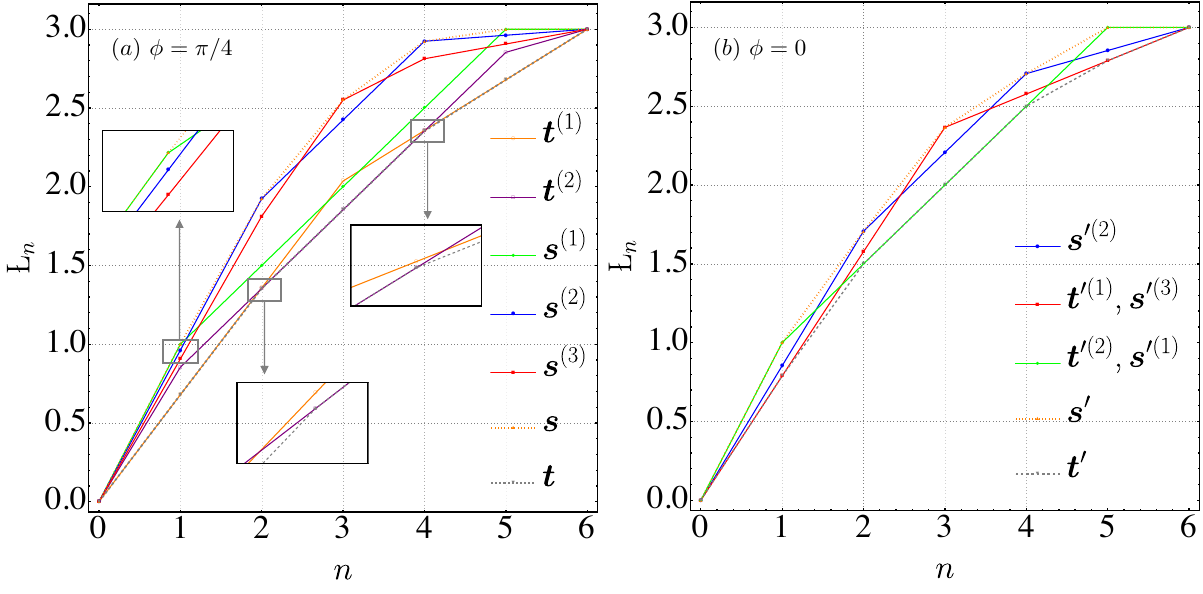}
\vspace{-2.5em}
\caption{The Lorenz curves of the UQCRs of three observables, where $L_{n}$ is the sum of the first $n$ components defined in equation (\ref{Lorenz curve}). By means of  $\boldsymbol{t}=\bigwedge_{n=0}^{5}\boldsymbol{t}^{(n)}$ and $\boldsymbol{s}=\bigvee_{n=0}^{5}\boldsymbol{s}^{(n)}$, the greatest lower bound and the least upper bound can be obtained, respectively. (a) The Lorenz curves of the infimum ($\boldsymbol{t}$) and supremum ($\boldsymbol{s}$) of the probability vectors $\boldsymbol{P}=\boldsymbol{p}^{(a)}\oplus\boldsymbol{p}^{(b)}\oplus\boldsymbol{p}^{(c)}$, where $\boldsymbol{t}\prec\boldsymbol{P}\prec\boldsymbol{s}$. The Lorenz curves of the $\boldsymbol{t}$ (the gray dashed antidiagonal line) and $\boldsymbol{s}$ (the orange dotted line) respectively give the greatest and the least possible envelope enclosing the curves of $\boldsymbol{P}$ for all quantum states; (b) The Lorenz curves of the infimum ($\boldsymbol{t}'$) and supremum ($\boldsymbol{s}'$) of the probability vectors $\boldsymbol{P}'=\boldsymbol{p}^{(d)}\oplus\boldsymbol{p}^{(b)}\oplus\boldsymbol{p}^{(c)}$, where $\boldsymbol{t}^{\prime}\prec\boldsymbol{P}^{\prime}\prec\boldsymbol{s}^{\prime}$. The Lorenz curves of the $\boldsymbol{t}^{\prime}$ (the gray dashed antidiagonal line) and $\boldsymbol{s}^{\prime}$ (the orange dotted line) respectively give the greatest and the least possible envelope enclosing the curves of $\boldsymbol{P}^{\prime}$ for all pure qubit states.}
\label{fig:3o}
\end{figure}

Similarly, the $\boldsymbol{t}^{(2)}$ and $\boldsymbol{t}^{(3)}$ can be obtained by one quantum state concurrently (See Figure \ref{fig:2t}(b)), i.e.,
\begin{equation}
\rho_{2(3)}=\frac{\mathds{1}}{2}+\frac{1}{2}\,\boldsymbol{r}_{2(3)}\cdot\boldsymbol{\sigma},\ \boldsymbol{r}_{2}=\boldsymbol{r}_{3}=\norm{\boldsymbol{r}}(1,0,0)\ .
\end{equation}
Since $\boldsymbol{t}^{(n)}$ also has the minimum summation of any $(6-n)$ components, then $\boldsymbol{t}^{(1)}=\boldsymbol{t}^{(5)}$, $\boldsymbol{t}^{(2)}=\boldsymbol{t}^{(4)}$.
Thus, we have
\begin{equation}
\begin{split}
\boldsymbol{t}^{(1)}
&=(\frac{1}{2}+\frac{1}{2}\norm{\boldsymbol{r}}\cos\vartheta,
\frac{1}{2}+\frac{1}{2}\norm{\boldsymbol{r}}\cos\vartheta,
\frac{1}{2}+\frac{1}{2}\norm{\boldsymbol{r}}\cos\vartheta,\\
&\quad\ \ \frac{1}{2}-\frac{1}{2}\norm{\boldsymbol{r}}\cos\vartheta,
\frac{1}{2}-\frac{1}{2}\norm{\boldsymbol{r}}\cos\vartheta,
\frac{1}{2}-\frac{1}{2}\norm{\boldsymbol{r}}\cos\vartheta)
=\boldsymbol{t}^{(5)}\ ,
\end{split}
\end{equation}
\begin{equation}
\boldsymbol{t}^{(2)}=
(\frac{1}{2}+\frac{1}{2}\norm{\boldsymbol{r}}\cos\phi,
\frac{1}{2},\frac{1}{2},\frac{1}{2},\frac{1}{2},
\frac{1}{2}-\frac{1}{2}\norm{\boldsymbol{r}}\cos\phi)=
\boldsymbol{t}^{(3)}=\boldsymbol{t}^{(4)}\ ,
\end{equation}
where $\cos\vartheta=\sqrt{\frac{2-\sqrt{2}}{6-\sqrt{2}}}$, $\phi=\pi/4$.
From \cref{lem:inf}, the greatest lower bound is
\begin{equation}
\label{eq:umb}
\begin{split}
\boldsymbol{t}
&=\bigwedge\lbrace\boldsymbol{t}^{(n)}|n=1,\cdots,5\rbrace\\
&=(\frac{1}{2}+\frac{1}{2}\norm{\boldsymbol{r}}\cos\vartheta,
\frac{1}{2}-\frac{1}{2}\norm{\boldsymbol{r}}\cos\vartheta+\frac{1}{2}\norm{\boldsymbol{r}}\cos\phi,\\
&\quad\ \ \frac{1}{2},\frac{1}{2},
\frac{1}{2}+\frac{1}{2}\norm{\boldsymbol{r}}\cos\vartheta-\frac{1}{2}\norm{\boldsymbol{r}}\cos\phi,
\frac{1}{2}-\frac{1}{2}\norm{\boldsymbol{r}}\cos\vartheta)\ .
\end{split}
\end{equation}
The vectors $\boldsymbol{t}^{(1)}$, $\boldsymbol{t}^{(2)}$ and $\boldsymbol{t}$ for pure states ($\norm{\boldsymbol{r}}=1$) are exhibited as shown in Figure \ref{fig:3o}(a). The Lorenz curve of each $\boldsymbol{P}=\boldsymbol{p}^{(a)}\oplus\boldsymbol{p}^{(b)}\oplus\boldsymbol{p}^{(c)}\in\mathcal{P}$ lies above the curve of $\boldsymbol{t}$, which illustrates the \cref{Thm:UCR}.

\textbf{Example 3}: In the case of $\phi=0$, considering three observables $D=\sigma_{x}$, $B=\sigma_{y}$, and $C=\sigma_{z}$ in a qubit system, of which the eigenbases are mutually unbiased. Then, from \cref{eq:umb}, the probability vectors $\boldsymbol{t}^{\prime (n)}$ are as follows:
\begin{equation}
\boldsymbol{t}^{\prime(1)}=
(\frac{1}{2}+\frac{1}{2\sqrt{3}}\norm{\boldsymbol{r}},
\frac{1}{2}+\frac{1}{2\sqrt{3}}\norm{\boldsymbol{r}},
\frac{1}{2}+\frac{1}{2\sqrt{3}}\norm{\boldsymbol{r}},
\frac{1}{2}-\frac{1}{2\sqrt{3}}\norm{\boldsymbol{r}},
\frac{1}{2}-\frac{1}{2\sqrt{3}}\norm{\boldsymbol{r}},
\frac{1}{2}-\frac{1}{2\sqrt{3}}\norm{\boldsymbol{r}})
=\boldsymbol{t}^{\prime(5)}\ ,
\end{equation}
\begin{equation}
\boldsymbol{t}^{\prime(2)}=
(\frac{1}{2}+\frac{\norm{\boldsymbol{r}}}{2},
\frac{1}{2},\frac{1}{2},\frac{1}{2},\frac{1}{2},
\frac{1}{2}-\frac{\norm{\boldsymbol{r}}}{2})
=\boldsymbol{t}^{\prime(3)}=\boldsymbol{t}^{\prime(4)}\ .
\end{equation}
Hence, by \cref{lem:inf}, the greatest lower bound can be calculated as
\begin{equation}
\boldsymbol{t}^{\prime}=
(\frac{1}{2}+\frac{1}{2\sqrt{3}}\norm{\boldsymbol{r}},
\frac{1}{2}+\frac{\norm{\boldsymbol{r}}}{2}-\frac{1}{2\sqrt{3}}\norm{\boldsymbol{r}},
\frac{1}{2},\frac{1}{2},
\frac{1}{2}-\frac{\norm{\boldsymbol{r}}}{2}+\frac{1}{2\sqrt{3}}\norm{\boldsymbol{r}},
\frac{1}{2}-\frac{1}{2\sqrt{3}}\norm{\boldsymbol{r}})\ ,
\end{equation}
where $\norm{\boldsymbol{r}}$ denotes the Euclidean norm of the Bloch vector $\boldsymbol{r}$. When $\norm{\boldsymbol{r}}=0$ (completely mixed states), the greatest lower bound $\boldsymbol{t}^{\prime}$ becomes trivial. For pure states, that is $\norm{\boldsymbol{r}}=1$, it turns out to be
\begin{equation}
\boldsymbol{t}^{\prime}=
(\frac{1}{2}+\frac{1}{2\sqrt{3}},1-\frac{1}{2\sqrt{3}},\frac{1}{2},
\frac{1}{2},\frac{1}{2\sqrt{3}},\frac{1}{2}-\frac{1}{2\sqrt{3}})\ .
\end{equation}
The Lorenz curves of $\boldsymbol{t}^{\prime}$ and $\boldsymbol{s}^{\prime}$ ($\norm{\boldsymbol{r}}=1$) are plotted in Figure \ref{fig:3o}(b). As indicated by \cref{Thm:UCR}, the Lorenz curve of each $\boldsymbol{P}'\in\mathcal{P}$ lies above the curve of $\boldsymbol{t}'$.

\begin{figure}[ht]
\centering
\includegraphics[width=0.8\linewidth]{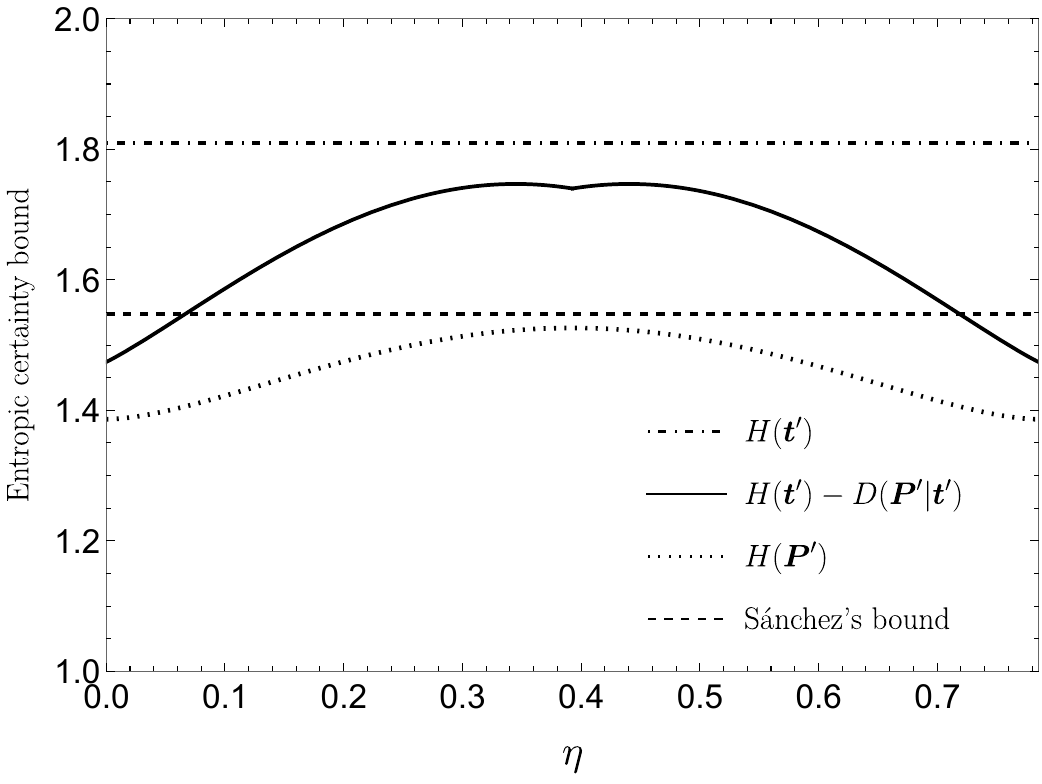}
\caption{Comparison of the entropic certainty relations' bounds. The dot-dashed curve and solid curve represent the bounds in Eq. (\ref{Shannon}) and Eq. (\ref{Shannon_tighter}), respectively. The dashed curve represents Sánchez's bound, and the Shannon entropy of $\boldsymbol{P^\prime}$ is represented by the dotted curve.}
\label{Fig-EQCR}
\end{figure}

For Pauli operators $\sigma_i$, $i=x$, $y$, $z$, Sánchez's entropic certainty relation is given by \(H(\boldsymbol{P^\prime}) \leqslant H_S = 1.54712\), which has long been considered optimal \cite{Canturk2021Optimal, Puchala2015Certainty}. We compare our entropic bounds in Eqs. (\ref{Shannon}, \ref{Shannon_tighter}) with Sánchez's bound for a class of states parametrized as $\ket{\psi} = \cos\eta\ket{0} + \sin\eta\ket{1}$, $\eta \in [0, \pi/4]$ (see Figure \ref{Fig-EQCR}). Results show that the bound in equation (\ref{Shannon_tighter}) is partially tighter than Sánchez's bound for this case, which has never been observed before.

Nevertheless, the probability vectors $\boldsymbol{p}_{\text{max}}$ that provide upper bounds for entropic quantum certainty relations in \cite{Sanchez1993Entropic,Sanchez1995Improved,Puchala2015Certainty,Canturk2021Optimal} maybe not generally comparable to the bound $\boldsymbol{t}$ in terms of majorization relation, i.e., neither $\boldsymbol{p}_{\text{max}}\prec\boldsymbol{t}$ nor $\boldsymbol{t}\prec\boldsymbol{p}_{\text{max}}$ holds. This would lead to paradoxical results for various entropic functions. In this sense, the majorization lattice theory provides a ``vector" measure for estimating quantum certainty and uncertainty, revealing the intrinsic structure of quantum measurement outcome distributions. In contrast, entropy and variance serve as scalar measures that unavoidably lose distribution information, such as the order of probability vector components. Consequently, different measures give rise to various bounds, uncovering diverse facets of quantum uncertainty. From this perspective, the structural property of the majorization lattice theory may account for the challenge of obtaining universal and optimal upper bounds for variances and entropies. More importantly, the examples show the universality of UQCR (\ref{glb_maj}) when applied to an arbitrary set of observables and mixed states, where entropic quantum certainty relations encounter difficulties. Computer programs may be required to efficiently compute the infimum of the majorization lattice, especially when dealing with a large number of observables or high-dimensional systems.

\section{Complementarity Relation for Quantum Coherence}
\label{CR_QC}
Quantum coherence, stemming from the superposition principle, is a fundamental feature of quantum physics. Exploring the interrelations between uncertainty and coherence is significant and has been discussed in various contexts \cite{LS17Q, SY21C}. We demonstrate that, using a vectorial measure, the uncertainty is identical to coherence for pure states. It is also shown that, by the majorization lattice properties, a complementarity relation for quantum coherence can be derived, which characterizes how much can the coherence with respect to one basis be, given a state has a high measure of coherence with respect to another basis. 

\begin{definition}[Coherent vector]
Let $\mathcal{B}=\lbrace\ket{i},\braket{i}{j}=\delta_{ij}\rbrace_{i=1}^{N}$, be an incoherent basis of $N$-dimensional Hilbert space. Accordingly, the coherence vector of the pure state $\ketbra{\psi}{\psi}$ is defined as 
\begin{equation}
\mu_{\mathcal{B}}(\ketbra{\psi}{\psi})\coloneqq(b_1,...,b_N)\ ,
\end{equation}
where $b_i=\norm{\braket{i}{\psi}}^2$, and $\sum_{i=1}^N b_i=1$. 
\end{definition}
Without loss of generality, we rearrange coherent vector $\mu_{\mathcal{B}}(\ketbra{\psi}{\psi})$ in descending order, i.e., $\mu_{\mathcal{B}}^{\downarrow}(\ketbra{\psi}{\psi})$ and omit the superscript $``\downarrow"$ hereafter. Thus, $\mu_{\mathcal{B}}^{\downarrow}(\ketbra{\psi}{\psi})\in\mathcal{P}^N$. 
 Given states $\ket{\psi}$ and $\ket{\phi}$, it is strightforward that if $\mu_{\mathcal{B}}^{\downarrow}(\ketbra{\psi}{\psi})\prec\mu_{\mathcal{B}}^{\downarrow}(\ketbra{\phi}{\phi})$, then $\ket{\psi}$ is more coherent than $\ket{\phi}$. Given an observable \( X = \sum_{i=1}^{N} x_i \ketbra{x_i}{x_i} \), its eigenstates form an incoherent basis \(\mathcal{X} = \{\ket{x_i}, \braket{x_i}{x_j} = \delta_{ij}\} \) in an \( N \)-dimensional Hilbert space. The coherence vector of the state \(\ket{\psi}\) is the probability distribution vector of the measurement \( X \), i.e., \(\boldsymbol{p}^{(x)} = \mu_{\mathcal{X}}(\ketbra{\psi}{\psi})\). In this context, the majorization quantum (un)certainty relation provides the complementarity relation for quantum coherence. Unfortunately, extending this to mixed states is not straightforward.

To define the coherence vector for a general state $\rho$, one can define the set of all the pure state decompositions of $\rho$ as
\begin{equation}
\mathcal{D}(\rho)=\Bigg\lbrace\lbrace q_{k},\ket{\psi_{k}}\rbrace_{k=1}^{L}\Big\vert\ \rho=\sum_{k=1}^{L}q_{k}\ket{\psi_{k}}\bra{\psi_{k}}\Bigg\rbrace\ .
\end{equation}
Here $L$ denotes the number of the pure states in the decomposition $\lbrace q_{k},\ket{\psi_{k}}\rbrace_{k=1}^{L}$, $\sum_k^L p_k=1$, $p_k\geqslant 0$ and $\ket{\psi_k}\in\mathcal{H}_N$ are normalized state vectors. Note that $\ket{\psi_k}$ are not necessarily orthogonal to each other. For clarification, $\mathcal{D}(\rho)$ is the set of pure state ensembles, which generate the same density operator $\rho$. It can be characterized by the classification theorem for ensembles \cite{MA2010Quantum}: $\rho=\sum_i p_i\ketbra{\psi_i}{\psi_i}=\sum_j q_j\ketbra{\phi_j}{\phi_j}$ for normalized states $\ket{\psi_i}$ and $\ket{\phi_j}$ if and only if for some unitary matrix $u_{ij}$
\begin{equation}
\sqrt{p_i}\ket{\psi_i}=\sum_j u_{ij}\sqrt{q_j}\ket{\phi_j}\ ,
\end{equation}
where $p_i$ and $q_j$ are probability distributions.
For the ensemble $\lbrace q_{k},\ket{\psi_{k}}\rbrace_{k=1}^{L}$, the mixed coherent vector is $\sum_{k=1}^{L}q_{k}\,\mu_{\mathcal{B}}(\ketbrad{\psi_k})$. Accordingly, we define the set of mixed coherence vectors for the pure decompositions of density operator $\rho$ as
\begin{equation}
\mathcal{U}(\rho)=\Bigg\lbrace\sum_{k=1}^{L}q_{k}\,\mu_{\mathcal{B}}(\ketbrad{\psi_k})\Big\vert\ \lbrace q_{k},\ket{\psi_{k}}\rbrace_{k=1}^{L}\in\mathcal{D}(\rho)\Bigg\rbrace\ .
\end{equation}
Hence, one can define the coherence vector for the general state $\rho$ as
\begin{equation}
\mu_{\mathcal{B}}(\rho)=\bigvee\mathcal{U}(\rho)\ ,
\end{equation}
which is the supremum of the set $\mathcal{U}(\rho)$ with respect of the majorization relation. It can be verified that the coherence vector $\mu_{\mathcal{B}}$ is a suitable vectorial quantifier of coherence \cite{Bosyk21Gen}.

\begin{figure}[ht]
\centering
\includegraphics[width=0.6\linewidth]{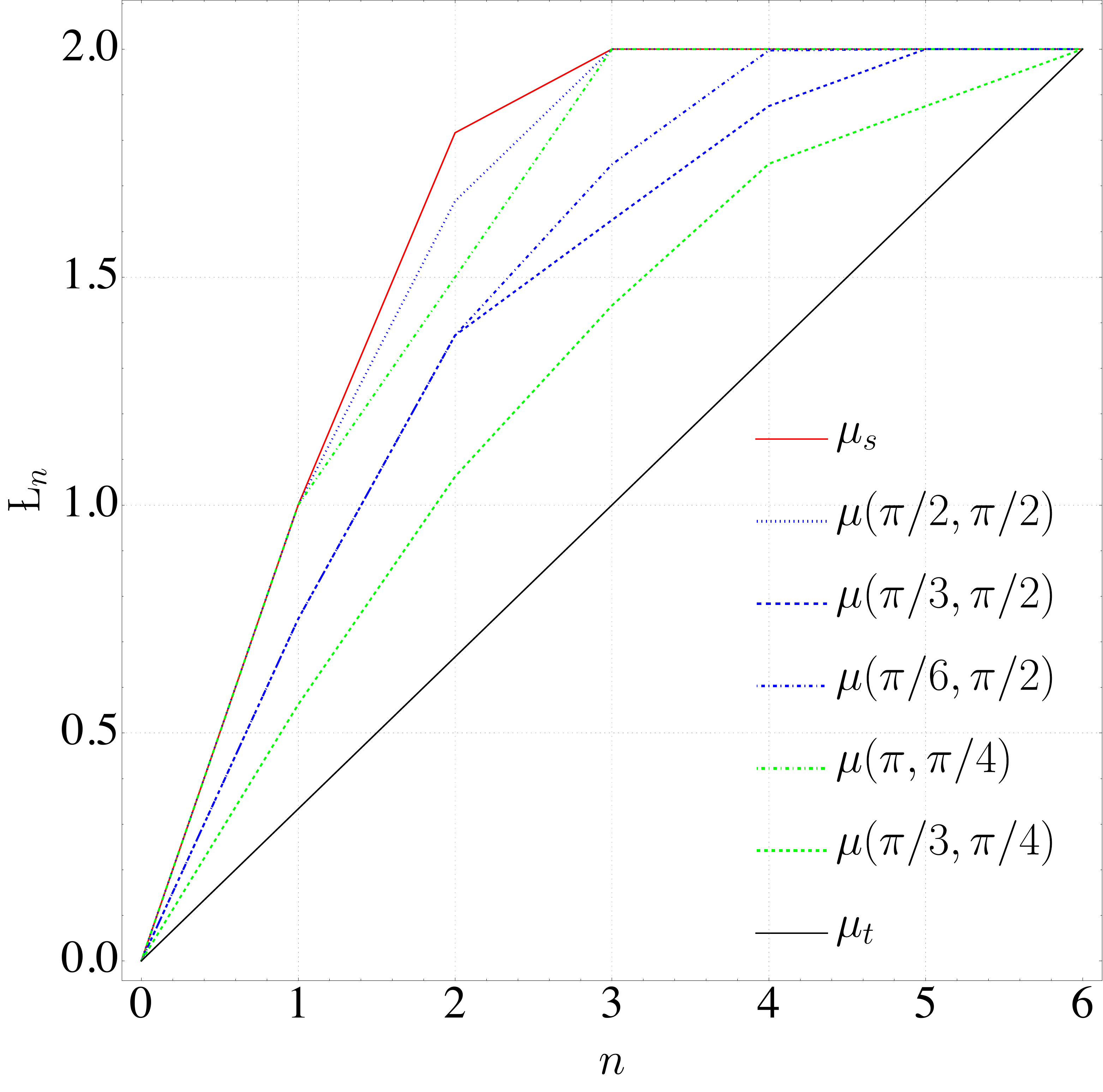}
\caption{The Lorenz curves of the bounds $\mu_s$ (red solid line), $\mu_t$ (black solid line) and the Lorenz curves of the coherent vector given by the random choices of five $\ket{\psi}$ (dotted, dashed and dot-dashed line), where any pair of $\theta$ and $\phi$, i.e., $(\theta,\phi)$ gives a quantum state.}
\label{fig:4c}
\end{figure}
	
Consider another incoherent basis $\mathcal{B}^\prime=\lbrace\ket{i^\prime}\rbrace_{i^\prime=1}^{N}$ and denote the coherence vector of as $\mu_{\mathcal{B}^\prime}(\rho)$. Then we have the coherence complementarity relation between different bases
\begin{equation}
\mu_t\prec\mu_{\mathcal{B}}\oplus\mu_{\mathcal{B}^\prime}\prec\mu_s\ ,
\end{equation}
where $\mu_t$ and $\mu_s$ symbolize the infimum and supremum of the $\mu_{\mathcal{B}}\oplus\mu_{\mathcal{B}^\prime}$, respectively. This complementarity relation is not limited to MUBs and applies to an arbitrary number of bases in principle. The Schur-concave functions allow us to quantify the complementarity relation for quantum coherence by scalar quantities. Here, we choose the Shannon entropy function to characterize the entropic complementarity relation as follows
\begin{equation}
H(\mu_s)\leqslant H(\mu_\mathcal{B})+H(\mu_{\mathcal{B}^{\prime}})\leqslant H(\mu_t)\ .
\end{equation}

We give an example to illustrate the coherence complementarity relation.\\
Example 4: We take two orthonormal bases $\mathcal{B}$ and $\mathcal{B}^\prime$ of the qutrit system 
\begin{equation}
\mathcal{B}=(\ket{a},\ket{b},\ket{c})=
\begin{pmatrix}
1&0&0\\
0&1&0\\
0&0&1
\end{pmatrix}\ ,
\end{equation}
\begin{equation}
\mathcal{B}^\prime=(\ket{\alpha},\ket{\beta},\ket{\lambda})=
\begin{pmatrix}
\frac{1}{\sqrt{3}}&\frac{1}{\sqrt{3}}&\frac{1}{\sqrt{3}}\\
\frac{1}{\sqrt{2}}&0&-\frac{1}{\sqrt{2}}\\
\frac{1}{\sqrt{6}}&-\sqrt\frac{2}{3}&\frac{1}{\sqrt{6}}
\end{pmatrix}\ ,
\end{equation}
and consider the state $\ket{\psi(\theta,\phi)}=\cos\theta\sin\phi\ket{a}+\sin\theta\sin\phi\ket{b}+\cos\phi\ket{c}$  as an example. For convenience, we denote $\mu(\theta,\phi)=\mu(\ketbra{\psi(\theta,\phi)}{\psi(\theta,\phi)})=\mu_{\mathcal{B}}(\ketbra{\psi(\theta,\phi)}{\psi(\theta,\phi)})\oplus\mu_{\mathcal{B}^\prime}(\ketbra{\psi(\theta,\phi)}{\psi(\theta,\phi)})$ here. Then, the supremum of $\mu(\theta,\phi)$, i.e., $\mu_s$ for any pure states reads
\begin{equation}
\mu_s=(1,\frac{\sqrt{6}}{3},1-\frac{\sqrt{6}}{3},0,0,0)\ .
\end{equation}
Using the fact that, for two orthogonal bases, one can always find a pure state that is unbiased with respect to both $\mathcal{B}$ and $\mathcal{B}^\prime$ \cite{KK14O}, hence, the infimum of $\mu(\ketbra{\psi(\theta,\phi)}{\psi(\theta,\phi)})$ is
\begin{equation}
\mu_t=(\frac{1}{3},\frac{1}{3},\frac{1}{3},\frac{1}{3},\frac{1}{3},\frac{1}{3})\ .
\end{equation}
The Lorenz curves of the bounds $\mu_s$, $\mu_t$ and the Lorenz curves of the coherent vectors given by the random choices of five $\ket{\psi(\theta,\phi)}$ are plotted in Figure \ref{fig:4c}, which shows that the direct sum of the coherence vectors corresponding to bases $\mathcal{B}$ and $\mathcal{B}^\prime$, respectively, for randomly selected quantum states are bounded by $\mu_s$ and $\mu_t$.

\section{Conclusions}
\label{Con}
We investigate in this work the quantum certainty relation by utilizing the majorization lattice theory. A universal quantum certainty relation is obtained, which is applicable to an arbitrary set of $M$ observables of any $N$-dimensional mixed and pure states. The UQCR unveils the intrinsic structural nature of quantum certainty and provides an optimal state-independent bound concerning the majorization relation. It is worth noting that the UQCR and the majorization uncertainty relation, encompassing all quantum states, provide optimal state-independent bounds and impose general constraints on the range of the joint probabilities, thereby characterizing the inherent incompatibility of the observables.

We exploit the UQCR to refine the entropic formulation certainty relation by the Shannon entropy function. The optimality of the UQCR can be exhibited explicitly by the Lorenz curves. Indeed, the direct-sum operation of vectors can be replaced by the direct-product operation, leading to distinct bounds. However, to explicitly determine the operation (direct-sum or -product) that may provide a tighter bound is not straightforward. It is worth mentioning that to find bipartite QCRs in the presence of quantum memory is primarily an open question. Finally, we explore the interrelations between uncertainty and coherence, and also derive a complementarity relation for quantum coherence between different bases. Notably, this relation applies to arbitrary bases, not just mutually unbiased bases.
\section*{Acknowledgements}
\noindent
This work was supported in part by National Natural Science Foundation of China(NSFC) under the Grants 12235008 and 12475087, and the University of Chinese Academy of Sciences.
\bibliographystyle{ref_style.bst}
\bibliography{References.bib}

\end{document}